\documentclass[a4paper,oldversion]{aa}
\usepackage{graphicx}
\usepackage{natbib}

\begin{document}

\title{Submillimeter and X-ray observations of an X Class flare}

\author{C.G. Gim\'enez de Castro \inst{1} 
\and    G. Trottet \inst{2}      
\and 	A. Silva-Valio \inst{1}    
\and 	S. Krucker \inst{3}
\and 	J.E.R. Costa \inst{4}
\and 	P. Kaufmann \inst{1,5}
\and 	E. Correia \inst{1,4}
\and 	H. Levato \inst{6}
}

\institute{Centro de R\'adio Astronomia e Astrof\'{\i}sica Mackenzie, R. da
      Consola\c{c}\~ao 896,  01302-907, S\~ao Paulo, SP, Brazil.
\and  LESIA, Observatoire de Paris, Section de Meudon, 92195 Meudon, France.
\and  Space Sciences Laboratory, University of California, Berkeley, USA.
\and  Instituto Nacional de Pesquisas Espaciais, S\~ao Jos\'e dos Campos,
      Brazil.
\and  Centro de Componentes Semicondutores, Universidade Estadual de Campinas, 
      Campinas, Brazil.
\and  Complejo Astron\'omico El Leoncito, CONICET, San Juan, Argentina.
}

\titlerunning{Submillimeter and X-ray observations of an X Class flare}
\authorrunning{Gim\'enez de Castro, Trottet, Silva-Valio et al}

\abstract{The GOES X1.5 class flare that occurred on August 30,2002 at 1327:30 UT is one
of the few events detected so far at submillimeter wavelengths. We present a
detailed analysis of this flare combining radio observations from 1.5 to 212
GHz (an upper limit of the flux is also provided at 405 GHz) and X-ray.
Although the observations of radio emission up to 212 GHz
indicates that relativistic electrons with energies of a few MeV where
accelerated, no significant hard X-ray emission was detected by RHESSI above
$\sim$ 250 keV. Images at 12--20 and 50--100 keV reveal a very compact, but resolved,
source of about $\sim 10\arcsec \ \times 10\arcsec$.  EUV TRACE images show
a multi-kernel structure suggesting a complex (multipolar) magnetic
topology.  During the peak time the radio spectrum shows an extended
flatness from $\sim 7$ to 35 GHz. Modeling the optically thin part of the
radio spectrum as gyrosynchrotron emission we obtained the electron
spectrum (spectral index $\delta$, instantaneous number of emitting
electrons). It is shown that in order to keep the expected X-ray emission
from the same emitting electrons below the RHESSI background at 250 keV, a
magnetic field above 500 G is necessary.  On the other hand, the electron
spectrum deduced from radio observations $\ge 50$ GHz is harder than that
deduced from $\sim 70 - 250$ keV X-ray data, meaning that there must exist
a breaking energy around a few hundred keV. During the decay of the
impulsive phase, a hardening of the X-ray spectrum is observed which is
interpreted as  a hardening of the electron distribution spectrum produced by the
diffusion due to Coulomb collisions of the trapped electrons in a medium
with an electron density of $n_e \sim 3 - 5 \ 10^{10} \
\mathrm{cm}^{-3}$. }

\maketitle

\section{Introduction}

During solar flares, a fraction of the released energy is used to accelerate
electrons with energies well above 1 MeV.  The interaction of these
particles with the magnetic field of the flaring region produces
gyrosynchrotron / synchrotron radiation observed at cm or smaller
wavelengths \citep[see e.g. ][ for
reviews]{Bastianetal:1998,PickVilmer:2008} and a bremsstrahlung continuum
caused by Coulomb collisions observed with X- and $\gamma$-ray
detectors. It was shown \citep[e.g.][]{Kunduetal:1994} that the electron
spectrum $N(E)$ determined by means of $\ge$ 30 GHz radio observations is
harder than that deduced from Hard X-ray (HXR) below a few hundred keV.
However for a few events, the electron spectra were found consistent with
spectra inferred from $\gamma$-ray continuum above $\sim$ 1 MeV
\citep{Trottetetal:1998,Trottetetal:2000}. Since radio emission above 30
GHz is produced mainly by electrons of a few MeV \citep[see
e.g. ][]{WhiteKundu:1992,Ramatyetal:1994}, these results have an impact on
acceleration mechanism models, which are still, an open question in solar
flare theory, and reinforces the need for good diagnostics of the $>$ 1
MeV particles.\\

Continuum X- and $\gamma$-ray detectors may observe photons from a few keV
up to tens of MeV, but have as a limitation the low sensitivity and / or
high background in the high energy range.  In the past three solar cycles
only a few tens of flares have been observed above 1 MeV. On the other
hand, radioastronomy at millimeter and submillimeter wavelengths is more
efficient than the $\gamma$-ray detectors. Routine solar flare observations
at 212 and 405 GHz started in March 2001 with the Solar Submillimeter
Telescope \citep[SST, ][]{Kaufmannetal:2001}, installed in the Argentinean
Andes.  A few flares were also observed at 210, 230, and 345 GHz with a
receiver array installed at the focus of the K\"oln Observatory for
Submillimeter and Millimeter Astronomy (KOSMA) telescope
\citep{Luthietal:2004a,Luthietal:2004b}. The first observations using such
instruments showed that the spectrum above 100 GHz is a continuation of the
cm-wavelength optically thin spectrum
\citep[e.g.][]{Trottetetal:2002,Luthietal:2004a} and extended the
diagnostic tools of radio observations to higher energy (a few tens of MeV)
electrons.  However, an unexpected upturn of the spectrum above 100 GHz was
reported for other $\ge$ M class events
\citep[e.g.][]{Kaufmannetal:2004,Luthietal:2004b,Cristianietal:2008}. The
physical processes responsible for the production of the spectrum upturn
are still unknown and are a subject of debate
\citep{KaufmannRaulin:2006,Silvaetal:2007,Trottetetal:2008}. \\

In this paper, we present a combined analysis of the impulsive phase of the
August 30, 2002, X class flare using RHESSI X-ray observations and
spatially unresolved radio data covering the range between 1.5 to 212 GHz
(and an upper limit for 405 GHz) obtained by different instruments. The
event has been analyzed by different authors. 
\cite{Karlickyetal:2004} related radio observations between 0.8 and
2.0 GHz and X-ray spectra and images from RHESSI. They found high-frequency
drifting structures between 1327:38 and 1327:50 UT with a global drift of
-25 MHz s$^{-1}$. The 10--20 keV X-ray sources show a north-east
displacement with a projected velocity of about 10 km s$^{-1}$, while the
29--44 keV emission is delayed by about 0.5 to 0.7~s after the radio
drifting structure.  Microwave observations of a short pulse during the
onset of this event was analyzed by \cite{GimenezdeCastroetal:2006} who found a
strikingly narrow spectrum that was explained as gyrosynchrotron emission
of accelerated electrons with a maximum energy (high energy cutoff) of about
250 keV. Another event with similar properties was qualitatively discussed by 
\cite{Luthietal:2004a}. In this work we extend the analysis of
\cite{GimenezdeCastroetal:2006} to the entire event. Moreover, we perform a quantitative 
analysis of the data which allows us to estimate the characteristics of the emitting electrons
(energy spectrum, total number) and of the flaring region (density, magnetic
field strength) that are necessary to account for the apparent discrepancy
between X-ray and radio observations.

\section{Instrumentation\label{sec:instrumentos}}

\begin{figure}
\centerline{
\resizebox{9cm}{!}{\includegraphics{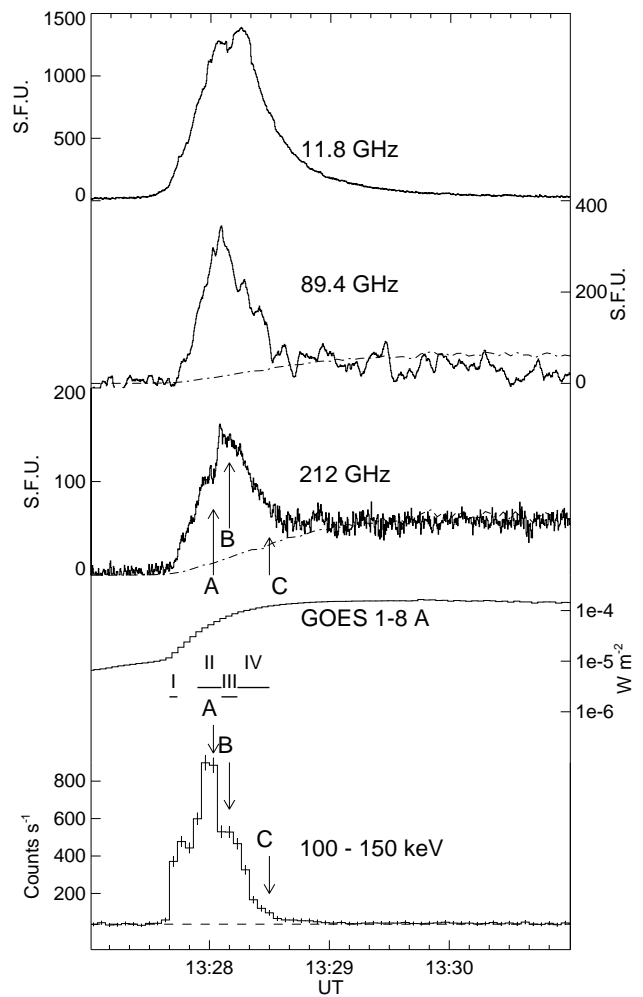}}}
\caption{Time evolution of the August 30, 2002 flare, at different
  radio frequencies and in selected SXR and HXR energy channels. At
  89.4 and 212 GHz the dashed curve represents the computed
  contribution of an isothermal source. A,B, and C indicate,
  respectively, characteristic time bins around the maximum of the HXR
  emission, the maximum of the 212 GHz radiation, and the decay
  phase of the burst. Horizontal bars denote the time intervals I through IV
  (see text).}
\label{fig:profiles}
\end{figure}

The hard X-ray (HXR) and radio data used in the present analysis of the
August 30, 2002 event were obtained with the NASA Reuven Ramaty High
Energy Solar Spectroscopic Imager (RHESSI), the Solar Submillimeter
Telescope (SST, installed in the Argentinean Andes), the nulling
interferometer and patrol telescopes of the University of Bern
(Switzerland), the Radio Solar Telescope Network (RSTN), and the Solar
Radio-polarimeter of the Radio Observatory of Itapetinga (ROI, Brazil).\\

RHESSI provides imaging and spectral HXR/$\gamma$-ray observations, with
high spatial ($\sim$ 2 arc sec) and spectral ($\sim$ 1 keV) resolution in
the $\sim$ 3 keV~--~17 MeV energy range \citep{Linetal:2002}.\\

The SST \citep{Kaufmannetal:2000} operates simultaneously at 212 and 405
GHz and with a time resolution of 1 ms for the present event.  The focal
system consists of four receivers at 212 GHz and two at 405 GHz. At 212 GHz
this produces a cluster of beams that, in principle, allows us to
determine the centroid of the emitting region whenever an event is detected
\citep[see ][ and references therein for details]{GimenezdeCastroetal:1999}.  During the
August 30, 2002 flare, SST was tracking NOAA region 10095, with one of the
two 405 GHz beams pointing at the active region.  At 212 GHz, the event
was observed with only one beam so that it was not possible to estimate the
centroid position of the emitting region.  The antenna temperatures have
been corrected for atmospheric attenuation (zenith optical depth
$\tau_{212} = 0.3$ nepers and $\tau_{405} = 1.35$ nepers) and converted to
flux density assuming that {\em (i)} the source is much smaller than the
beam size and {\em (ii)} there is no important main-lobe gain correction
due to a possible offset pointing.  Since the HPBW of the beams are
respectively $\sim$ 4$\arcmin$ and 2$\arcmin$ at 212 and 405 GHz,
hypothesis {\em (i)} is justified here, because this event is very compact
in the HXR domain (see Sect. \ref{subsubsec:hxrimages}). On the other hand,
the projected position of the beam that observed the burst in the sky is
separated by less than 30$\arcsec$ from the HXR emitting region observed by
RHESSI. As this is comparable to the absolute position uncertainty of the
SST antenna, hypothesis {\em (ii)} is also justified. It should be
emphasized that a misalignment of 30$\arcsec$ produces a main-lobe gain
correction of less than 5\% at 212 GHz (HPBW $\sim$ 4$\arcmin$).\\

The two-element nulling interferometer of the University of Bern provides
total flux measurements at 89.4 GHz with a sensitivity of $\sim$ 35
s.f.u. (1 s.f.u. = 10$^{-22}$ W m$^{-2}$ Hz$^{-1}$) and a time resolution
of 31 ms \citep{Luthietal:2004a}. Total flux densities at 11.8, 19.6, 35,
and 50 GHz were recorded by the patrol telescopes at Bumishus (Switzerland)
with a time resolution of 100 ms. Total flux density measurements made at
1.415, 2.695, 4.995, 8.8, and 15.4 GHz by the RSTN with a time resolution
of 1 s, and at 7 GHz by the Solar Radio-polarimeter of the Radio
Observatory of Itapetinga with 20 ms time resolution have also been used.\\

\begin{figure}
\centerline{
\resizebox{9cm}{!}{\includegraphics{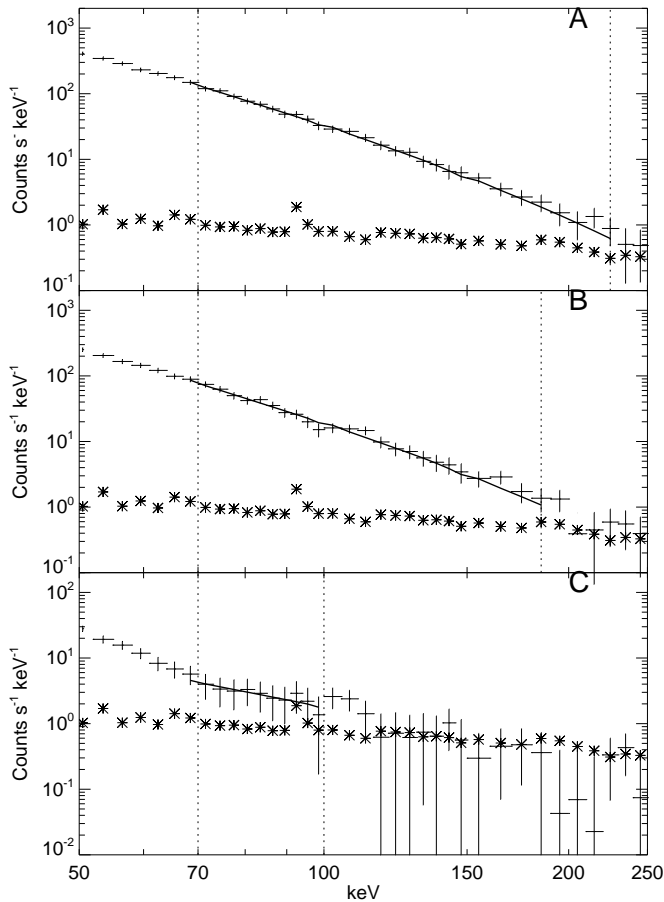}}}
\caption{Background subtracted count rate spectra measured by RHESSI
  (error bars) during time bins A,B, and C marked in
  Fig. \ref{fig:profiles}.  The asterisks show the background
  spectrum. The continuous lines show the best fit models (see
  text). The vertical dashed lines indicate the fitted energy range.}
\label{fig:spex}
\end{figure}

\section{Observations and data analysis\label{sec:observaciones}}

The August 30, 2002 HXR and radio event starting at $\sim$ 1327:30 UT is
associated with a GOES X1.5 SXR burst and, with a surprisingly small,
H$\alpha$ sub-flare which occurred in NOAA Active Region 10095 (N15
E74). This event was observed by RHESSI up to $\sim$ 250 keV around its
maximum and in a large part of the radio spectrum ranging from
submillimeter to decameter and longer wavelengths.  Figure
\ref{fig:profiles} displays the time profile of the event in the 100 -- 150
keV HXR band, in the SXR 1--8 \AA \ channel; it also shows the total flux
densities at 11.8, 89.4, and 212 GHz. The flare comprises an impulsive
phase that starts at $\sim$ 1327:40 UT in the 100 -- 150 keV band and which
lasts for about 60 s.  \\

\subsection{Flux and spectra\label{subsec:flujos}}
\subsubsection{Hard X-rays\label{subsubsec:flujoHXR}}

\begin{figure}
\centerline{
\resizebox{9cm}{!}{\includegraphics{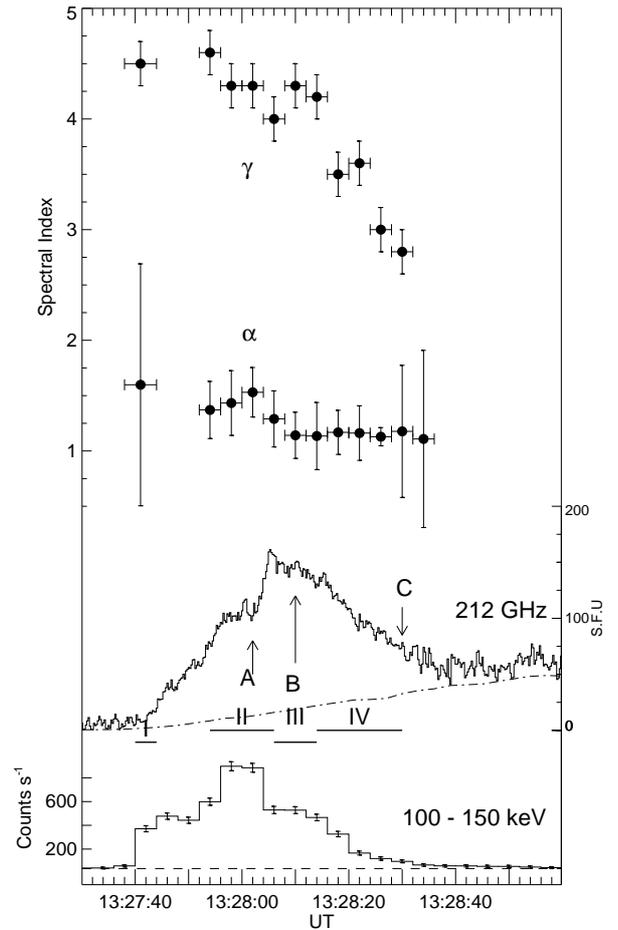}}}
\caption{From top to bottom: time evolution of the HXR spectral index
  $\gamma$ and of the radio spectral index $\alpha$ (dots), of the radio
  flux density at 212 GHz (the dot-dashed curve represents the computed
  contribution of an isothermal source), and of the 100 -- 150 keV
  emission.  Horizontal bars  show the time intervals I through IV (see text).}
\label{fig:alpha}
\end{figure}

Spectral analysis of RHESSI data was performed between 1327:28 and
1328:36 UT.  Count rate spectra were accumulated between 1327:38 and
1327:44 UT (before the thick shutter came in) and in all 4 second intervals
between 1327:52 and 1328:32 UT using front detectors 1, 3--6, 8 and 9.  We
applied pile up and decimation corrections.  Each spectra consists of 77
energy bands between 3 and 250 keV.  For each interval, spectral fitting was
carried out for energies ranging from 40 keV to the highest energy where
count rates in excess of 2~$\sigma$ above background are measured.  It is
found that the count rate spectra could be reasonably represented by
considering either a single power law or a double power-law.  In the latter
case, the break energy lies around 70 keV.  Since we are interested in the
non-thermal X-ray emission, we have restricted the analysis to the energies
above 70 keV using a single power-law of spectral index $\gamma$ for the
trial photon spectra.  Fig \ref{fig:spex} displays examples of the fitted
spectra for the time bins labeled A, B, and C in Fig. \ref{fig:profiles}.
The time evolution of $\gamma$ is shown in Fig. \ref{fig:alpha}.\\

\subsubsection{Radio \label{subsect:fluxuw}}

\begin{figure}
\centerline{
\resizebox{9cm}{!}{\includegraphics{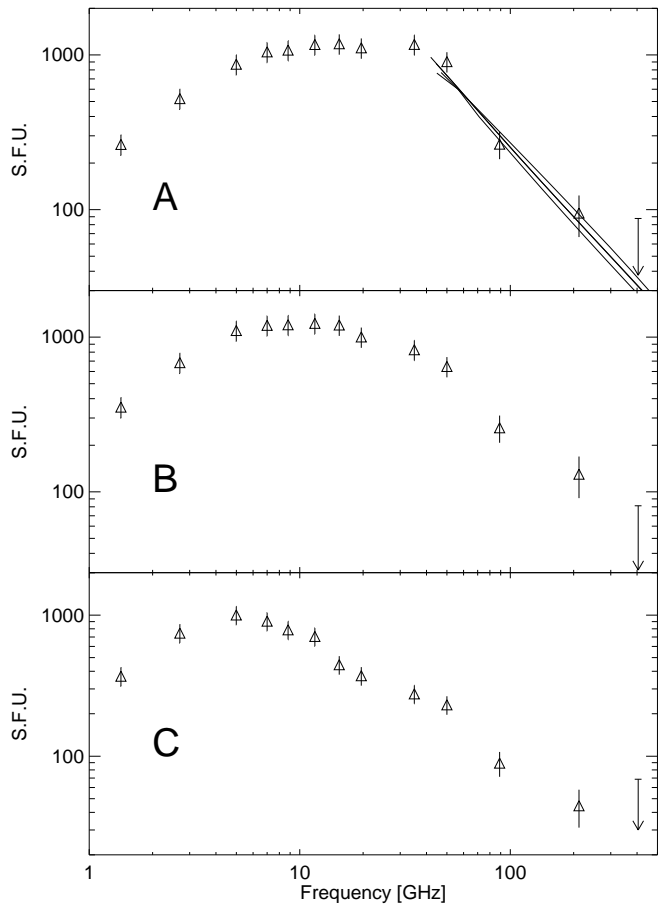}}}
\caption{The radio spectrum at instants labeled A,B, and C.  (see Figure
\ref{fig:profiles}) Solid curves represent the homogeneous
gyrosynchrotron solution fits discussed in Sect. \ref{subsect:relation}.}
\label{fig:mwspectra}
\end{figure}

At frequencies above 20 GHz there is a time extended emission that lasts
for tens of minutes after the impulsive phase (see Fig \ref{fig:profiles}
at 89.4 and 212 GHz). This gradual component is most likely of thermal
origin because it seems to follow the SXR emission from GOES and has no HXR
counterpart. The comparison of the $>$ 20 GHz time evolution with that of
the 1 -- 8 \AA \ SXR indicates that this thermal component may have started
at the beginning of the impulsive phase and therefore should be subtracted
from the total flux densities in order to estimate the non thermal emission
of the radio burst. For this we consider that the thermal emission arises
from the hot thermal source that produces the SXR emission observed by
GOES. We computed the free-free flux density of an isothermal source with
temperature and emission  measure (EM) derived from GOES 8 observations.  A
source size of 60\arcsec\ provides a reasonable agreement with the observations
as illustrated in Fig. \ref{fig:profiles} (dot-dash lines at 89.4 and 212
GHz curves).  Fig. \ref{fig:mwspectra} shows the non-thermal radio spectrum
for the three time bins marked A, B and C in Fig. \ref{fig:profiles} that
correspond, respectively, to the maximum of the 100--150 keV emission, to the
maximum of the 212 GHz, and to the decay of the impulsive phase. The main
characteristics of the radio spectra can be summarized as follows:

\begin{description}
\item[From 1 to 7 GHz:] the spectrum increases with frequency and can be
  represented by a power law with spectral index $\sim 1$. 
\item[From 7 to 35 GHz:] there is a plateau observed during the
  impulsive phase (Fig. \ref{fig:mwspectra}, A and B). During the
  decay of the impulsive phase the spectrum gradually evolves exhibiting a 
  rather well defined turnover frequency around 5 GHz (Fig. \ref{fig:mwspectra}, C).
\item[Above 50 GHz:] the flux density is roughly proportional to
  $\nu^{-\alpha}$. Fig. \ref{fig:alpha} shows the time evolution of 
  $\alpha$. \\
\end{description}

\subsection{HXR Images\label{subsubsec:hxrimages}}

\begin{figure}
\centerline{
\resizebox{9cm}{!}{\includegraphics{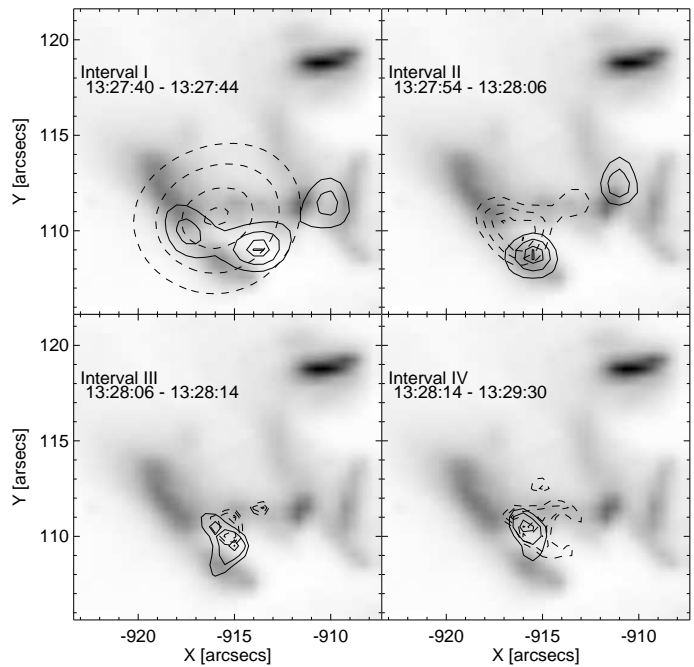}}}
\caption{RHESSI emitting regions of 50 -- 100 keV (solid contours) and
  of 12 - 20 keV (dashed contours) for different time intervals
  superimposed on a negative 195 \AA\ image taken by TRACE at 1327:31
  UT. Contour levels are: 50, 70, 90, and 99\% of the image maximum.}
\label{fig:rhessi}
\end{figure}

Figure \ref{fig:rhessi} displays RHESSI contours in the 12--20 keV and
50--100 keV bands overlaid on a 195 \AA\ TRACE image taken at 1327:31 UT
for the four intervals marked by horizontal bars in
Fig. \ref{fig:profiles}. Intervals I and II cover the first two 100 -- 150
keV peaks, interval III spreads over the maximum of the 212 GHz, while
interval IV extends over the HXR decay phase.  RHESSI images were obtained
by applying the PIXON algorithm \citep{Metcalfetal:1996} and by using front
detector segments 1 to 6.  Detector 2 was not used for the 12 -- 20 keV
images.  The TRACE image was taken close to the onset of the radio and HXR
impulsive peaks. The figure shows that the impulsive phase of the flare was
triggered within a complex pattern of bright EUV features in a compact region
(10\arcsec $\times$ 10\arcsec).  Unfortunately, subsequent 195 \AA\ TRACE
images are saturated until 1350 UT. During intervals I and II, the 50--100
keV emission (solid contours) arises from two compact regions ($\sim$
3--7\arcsec) overlaying bright EUV structures. During the rest of the event
(intervals III to IV), only the southeast source is observed at 50 -- 100
keV.  During the whole event the lower energy HXR (12 -- 20 keV, dashed
contours) arise predominantly from a single source close to the southeast
of the 50 -- 100 keV emitting region. However, the interpretation of the HXR 
images at 50 keV for interval III and IV is inconclusive as up to half of the 
counts are due to pile up of thermal photons.  To correct pile up in images is 
very difficult as the corrections should be done in the modulated light curves. 
Currently, the RHESSI software does not provide pile up correction in images.

\section{Discussion\label{sec:discusion}}

EUV observations close to the onset of the impulsive phase of the August
30, 2002 event reveal that this flare arose from a compact region with a
multi kernel structure that suggests a complex (multipolar) magnetic field
topology.  The HXR emitting sources are observed in association with
different EUV bright structures, suggesting that they are located at the
footpoints of different magnetic loops (see Fig. \ref{fig:rhessi}). The
compactness of this flare is further supported by the fact that such an
X class event produced only an H$\alpha$ sub-flare. In the following, we
discuss some peculiarities of this HXR and radio event:
\begin{itemize}
\item the unusually flat radio spectrum between 7 and 35 GHz
  during most of the impulsive phase (intervals I to III on
  Fig. \ref{fig:profiles}),
\item the lack of  significant
  $\gamma$ ray emission detected by RHESSI in the MeV domain,  
  although $\ge$ 90 GHz radio data indicate that relativistic
  electrons were produced during the flare,
\item the strong hardening of the $>$ 70 keV HXR spectrum during the
  decay of the impulsive phase.
\end{itemize}

Fig. \ref{fig:alpha} (top) presents the time evolution of the HXR spectral
index $\gamma$ obtained assuming that the photon distribution is a single
power law from 70 keV to the highest energy where count rates are above the
background (see Sect. \ref{subsubsec:flujoHXR}). During intervals I to III,
the spectral index does not vary significantly, remaining around $\gamma =
4.3 \pm 0.3$, however during interval IV there is a spectral  hardening.
We consider that intervals I, II, and III correspond to three successive
different particle injections:
\begin{itemize}
\item during intervals I and II, the 100 -- 150 keV emission is
  consistent with thick-target emission because it arises
  predominantly from footpoint sources. Furthermore there is no long-term coronal 
  trapping because the HXR peaks occur simultaneously at
  all energies within the RHESSI time resolution (4 s). Thus, the time
  evolution of the HXR emission mimics that of the electron injection
  into the thick-target region.
\item Taking into account the above arguments, interval III
  corresponds to a third injection which is reflected by the shoulder
  in the 100 -- 150 keV time profile range and corresponds to the
  maximum of the 212 GHz emission. 
\item The hardening of the HXR spectrum during interval IV may be indicative 
  of some trapping.
\end{itemize}

\subsection{The flat radio spectrum}
\label{sec:flatness}

It is well documented that cm-mm emission comes from gyrosynchrotron radiation of
energetic relativistic electrons propagating in the magnetic structures
\citep[e.g. ][ and references therein]{PickVilmer:2008}. The radio spectra
of the present event are indeed reminiscent of this emission process.  In a
uniform magnetic field, the emission would lead to a spectral index larger
than 2.5 in the optically thick region and a rather well defined turnover
frequency. However, the spectral index for the frequency range 1 -- 7 GHz
(optically thick emission) remains around or below 1 during the whole
event.  And until the decay of the impulsive phase (interval IV) the radio
spectrum is flat between 7 and 35 GHz (see Fig. \ref{fig:mwspectra}). Both
these characteristics are indications that radio emitting electrons
propagate in a highly inhomogeneous magnetic field
\citep[e.g.][]{Dulk:1985,Kleinetal:1986,Leeetal:1994}.  This inhomogeneous
magnetic field interpretation is consistent with the complex magnetic
structure revealed by TRACE and RHESSI images.  During the decay of the
impulsive phase, a well defined turnover frequency gradually becomes better
defined. At the same time, the X-ray emission arises from a single source,
so that the observed radio spectrum is closer to that expected from a
homogeneous source. \\

\citet{RamatyPetrosian:1972} explained flat microwave spectra as observed
by \citet{HachenbergWallis1961} by including the free-free absorption of a cold
medium uniformly mixed in the homogeneous gyrosynchrotron source region.
Indeed: (i) at low frequencies, where both gyrosynchrotron and free-free
opacities are $>$~1, the radio flux increases with frequency; (ii) at
frequencies for which the gyrosynchrotron opacity is $<$ 1 while the free-free
opacity remains $>1$, a plateau is observed because while the
gyrosynchrotron emission starts to decrease the free-free emission is still
increasing; (iii) at higher frequencies both emissions are optically thin and
the radio spectrum decreases with frequency.
The observation of simultaneous brightenings and line broadening of
hot ($\sim10^7$ K) and cold ($\sim 10^4$ K) plasmas during a solar limb
flare \citep{Kliemetal:2002} provides some support to the Ramaty \&
Petrosian hypothesis.  In addition to magnetic field inhomogeneities,
free-free absorption may also contribute to provide the observed flat radio
spectrum. In that case, as the flare evolves, the free-free opacity should
decrease in order to allow lower frequency radiation to become optically
thin.  Since free-free opacity is proportional to $n_p^2 T_p^{-3/2}$ ($n_p,
\ T_p$ medium density and temperature, respectively, and assuming $n_e =
n_p$), this would imply either a decrease of the density, or an increase of
the temperature, or an increase of both. In the latter case the temperature
should increase faster than the density.\\

Above 50 GHz (optically thin emission), the radiation,  which is mostly emitted
by highly relativistic electrons,   is not affected by
the medium, and the spectral radio index $\alpha$ is only related to the
index of the instantaneous electron distribution $\delta$. For the present
event $\alpha$ remains between 1.1 and 1.3 (Fig \ref{fig:alpha}).  Considering the
ultra-relativistic case as a gross approximation, this leads to $\delta =
2\alpha + 1 = 3.2 - 3.6$ \citep{Dulk:1985}.

\subsection{HXR spectral hardening during the decay of the impulsive phase}
\label{subsect:trapping}

During interval IV the HXR spectral index $\gamma$ decreases from 4.1 to 2.8
(Fig. \ref{fig:alpha}). This provides some indication that a  significant
amount of HXR was produced by trapped electrons, the diffusion rate being
governed by Coulomb collisions.  For the weak-diffusion limit case, the
trapping time $t^{trap} \simeq t^{def}$, with $t^{def}$ the characteristic
deflection time, which is given by \cite{Trubnikov:1965} and
\cite{MelroseBrown:1976}
\begin{equation}
t^{def} \simeq \frac{E}{2}\left ( \frac{dE}{dt} \right
)^{-1}_{coll} \ ,
\end{equation}
where $(dE/dt)_{coll}$, the electron energy loss rate due to Coulomb
collisions, is approximated by \citep{BaiRamaty:1979}
\begin{equation}
\left ( \frac{dE}{dt} \right )_{coll} = \left \{ 
\begin{array}{l@{\quad}l}
-4.9 \ 10^{-9} \ E^{-0.5} \ n_e & E \le 160 \ \mathrm{keV}  \\
-3.8 \ 10^{-10} \ n_e & E > 160 \ \mathrm{keV}  \\
\end{array}
\right .
\end{equation}
with $E$ the electron energy in keV and $n_e$ the medium electron
density in $cm^{-3}$.\\

We consider that the injection of electrons in the HXR emitting region
stopped at 1328:14 UT because it is the last time bin where $\gamma$
remains almost constant (see Fig. \ref{fig:alpha}). Under these conditions
the time evolution of the distribution of the trapped electrons is
approximately given by \citep{Aschwanden:1998}
\begin{equation}
N(E,t) \propto E^{-\delta} \exp{\left (-\frac{t}{t^{def}} \right )} \ ,
\end{equation}
with $\delta$ the electron index at the end of the injection. The
photon flux at energy $\epsilon$ as a function of time $t$ may be
written as \citep[see e.g.][]{Brown:1971}
\begin{equation}
F(\epsilon,t) = n_e \int_\epsilon^\infty  \sigma(\epsilon,E)   v(E) N(E,t) dE \ ,
\label{eq:brown}
\end{equation}
where $v(E)$ is the electron velocity corresponding to energy $E$,
$\sigma(\epsilon,E)$ is the bremsstrahlung differential cross-section per
unit photon energy $\epsilon$ for an electron of energy $E$. For $\sigma(E)$
we adopted the Bethe-Heitler electron -- proton bremsstrahlung
cross-section.  It should be noted that the photon flux produced by
precipitated electrons has a time evolution as $F(\epsilon,t)$
\citep{MelroseBrown:1976}.  Therefore, for a given $\delta$, the time
evolution of the photon flux depends only on $n_e$. By using equation
\ref{eq:brown} we computed the expected photon flux $F(\epsilon,t)$ and
compared its time evolution at different energies $\epsilon$ with that
observed by RHESSI during the decaying of the impulsive phase.  Comparing
the expected and observed time profiles at 70, 100, and 150 keV, we
obtained reasonable agreement for $n_e \sim 3 - 5 \ 10^{10} \
\mathrm{cm}^{-3}$ when $\delta$ ranges between 3.5 and 5, which correspond
respectively to thin and thick target limits for $\gamma\sim 4$. 
\cite{Kruckeretal:2008} find similar electron densities for coronal $\gamma$-ray 
emission during flares.\\

Spectral hardening has been reported during the impulsive phase of long duration
GOES X class flares and associated with non thermal footpoint bremsstrahlung 
\citep{Qiuetal:2004,Saldanhaetal:2008}. \cite{GrigisBenz:2008} analyzed 
the spectral hardening during the gradual phase of great flares and 
concluded that the cause is the continuing acceleration with longer trapping in 
the accelerator before escape. \cite{Kiplinger:1995} has shown that the hardening 
is associated with SEP events.

\subsection{Relationship between HXR and radio emission}
\label{subsect:relation}

The absence of HXR emission $>$ 250 keV while we observe radio emission
above 50 GHz can be used to constrain the high energy electron
distribution, the magnetic field, and the trapping time in the radio
emitting region. For that, we consider that non thermal electrons are
injected in coronal loops. The radio emission is produced in the coronal
portion of these loops where they become partially trapped while
precipitating electrons produce the HXR radiation by thick target
interaction at the loop footpoints. Since no spatially resolved radio data
are available for this event and the optically thin part of the radio
spectrum does not depend on the structure details of the medium, we 
used a homogeneous model to derive the mean parameters of the radio source
and emitting electrons.  For that, we computed the radio optically thin
emission by using the numerical code for a gyrosynchrotron source with a
homogeneous ambient density and magnetic field and  an isotropic electron
distribution developed by \cite{ramaty:1969} and corrected by
\cite{Ramatyetal:1994}\footnote{See \tt
http://lheawww.gsfc.nasa.gov/users/ramaty}. The instantaneous electron
distribution in the radio source was taken as $N(E) = K E^{-\delta}$, where
$K$ is the number of electrons per MeV at 1 MeV and the energy $E$ is in
MeV. The angle between the observer and the magnetic field (view-angle) was
set to 84$^\circ$ (the maximum allowed value in Ramaty's solution) in order
to obtain the lower limit of the total number of electrons necessary to
produce a given radio spectrum for a given magnetic field strength. For a
view-angle of 45$^\circ$, as is usually assumed, the computed number of
electrons is roughly twice that obtained for 84$^\circ$.  Table \ref{tbl:mwspec}
displays the values of the instantaneous total number of electrons above 25
keV $N(> 25 \mathrm{keV})$ and $K$ obtained for different values of the
magnetic field at the maximum of the 100 -- 150 keV HXR (time bin A) . $N(>
25 \mathrm{keV})$ was computed for $\delta = 3.4$ which provides the best
fit to the radio data. This is in agreement with the value of $\delta$ inferred 
from the slope of the optically thin part of the radio spectrum $\alpha$
(see sect. \ref{sec:flatness})\\

\begin{table}
\caption{Derived characteristic of radio radiating electrons for
  different magnetic field strengths and view angle equal to
  84$^\circ$. \label{tbl:mwspec}}
\begin{tabular}{ccc}
\bf B    &  N($> 25$ keV) $\times 10^{33}$ & $K \times 10^{29}$\\
(Gauss)  &                                  & ($e^-$ MeV$^{-1}$)\\
\hline
 500 &  38.1 &  76.0 \\
 750 &  15.0 &  30.0 \\
1000 &   7.5 &  15.0 \\
1600 &   1.5 &   3.0 \\
\hline
\end{tabular}
\end{table}

\begin{figure}
\centerline{
\resizebox{6cm}{!}{\includegraphics{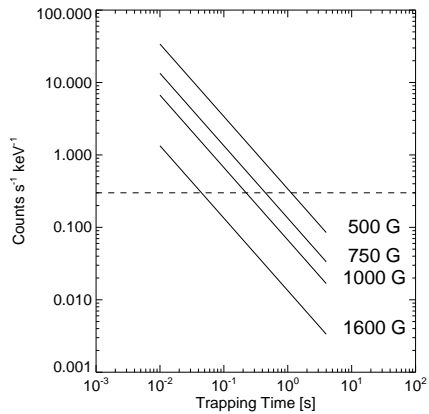}}}
\caption{Expected RHESSI count rates at 250 keV as a function of the
emitting electron trapping time for different magnetic field strengths
and view angle equal to 84$^\circ$.  Each solid line represents a
different solution determined by a different magnetic field. The
dashed horizontal line represents the measured background count rate at 250
keV.}
\label{fig:thick}
\end{figure}

The mean electron flux $\dot N(E)$ entering the thick target HXR
source is $\dot N(E) \simeq N(E)/t^{trap}$, where $t^{trap}$ is the
time spent by the electrons in the radio source. The thick target
photon emission from these precipitating electrons was calculated by
using a numerical code that takes into account both electron-proton
and electron-electron collisions. The photon spectrum is then
convolved with the RHESSI response matrix to get the corresponding
count rate spectrum. Figure \ref{fig:profiles} shows that within the RHESSI
time resolution (4 s), the 89.4 GHz, 212 GHz and 100-150 keV count rates 
show simultaneous peaks during intervals I to III.  The electron trapping 
time $t^{trap}$ can thus be considered independent of energy as a first 
approximation.  This suggest that during intervals I to III the precipitation 
rate is more likely governed by wave-particle interactions (turbulent trapping) 
than by Coulomb collisions \citep[e.g. ][]{Vilmer:1987}.
The computations were then carried out for $10^{-2}
\le t^{trap} \le 4 \ \mathrm{s}$, the lower and upper limits
correspond respectively to free streaming of the electrons in a
compact loop ($\sim 10$\arcsec), and to the RHESSI time resolution.
Figure \ref{fig:thick} displays the RHESSI expected count rates in the
250 -- 265 keV band as a function of $t^{trap}$ for different values
of the magnetic field. The dashed horizontal line corresponds to the
RHESSI count rate in this channel, the highest energy where this event
was detected.  We conclude that the mean magnetic field strength
should be greater than about 500 G to keep the thick target photon flux
expected from radio emitting electrons with trapping times smaller
than 1 s below the detection limit of RHESSI.\\

It has been shown that the thick target spectral index $\gamma$ is bounded
between $\delta - 1.5 \le \gamma \le \delta - 1$, the lower limit
corresponding to turbulent trapping of electrons with energies above a few
100 keV, whereas the higher limit is set by free propagating electrons with
energies below a few 100 keV \citep[see e.g.][ and references
therein.]{Trottetetal:1998}. Therefore if $\delta = 3.4$, the HXR emission
of these electrons should have an index between $1.9 \le \gamma \le 2.4$
which is significantly smaller than the observed values, $\sim 4.3$, during
intervals I to III (see Fig. \ref{fig:alpha}). This is in accordance with
previous works that show that radio high frequency emission is generated by
electrons with energies above $\sim$ 500 keV with an electron index harder
than the $<$ 500 keV electrons
\citep{WhiteKundu:1992,Kunduetal:1994,Trottetetal:1998,Silvaetal:2000}. \\

\section{Summary\label{sec:conclusiones}}

In this paper we have analyzed X-ray observations from RHESSI and radio
data obtained at submillimeter wavelengths by the Solar Submillimeter
Telescope (SST) of the X1.5 event that occurred in Active Region 10095 on August
30, 2002, at 1327:30 UT, complemented with radio observations from 1.5 to
89.4 GHz from other instruments.  EUV images from TRACE provided
information about the source emitting region. \\

The radio spectrum above 100 GHz is the continuation of the optically thin
microwave spectrum, therefore does not belong to the so called {\em THz}
bursts \citep{Kaufmannetal:2004}, although it is an X Class flare.  We
summarize below our main findings:
\begin{itemize}
\item The magnetic structure of the flare  is complex and
      highly inhomogeneous. This is revealed 
      by the 50 -- 100 keV and EUV images.  Such an inhomogeneous source
      may produce the flatness in the radio spectrum observed between 7 
      and 35 GHz, although we do not discard the free-free absorption.
\item The electron spectrum $N(E)$  above 1 MeV is harder than that at energies 
      below a few hundred keV.
\item Modeling simultaneously the expected RHESSI count rate and the expected 
      gyrosynchrotron emission, we obtain 500 G as a  lower limit for 
      the mean magnetic field of the flaring region. 
\item The time evolution of the spectral index deduced from X-ray
      observations at the end of interval III suggests that trapped
      electrons are diffused by Coulomb collisions.  This leads to a mean ambient 
      electron density of $3 - 5 \ 10^{10}$ cm$^{-3}$, typical of the low 
      corona / upper chromosphere and is compatible with previous results 
      \citep{Kruckeretal:2008} and with the small size of 
      the EUV pattern observed by TRACE, which also suggests that the flaring 
      region does not extend high in the corona.
\end{itemize}

Finally it should be emphasized that radio observation in the sub-THz domain 
provide  a unique tool to constrain acceleration model because they constitute 
a more sensitive diagnostic of ultra-relativistic electrons than present 
HXR/$\gamma$-ray measurements. 

\begin{acknowledgements}
This research was partially supported by Brazil Agencies FAPESP, CNPq and
Mackpesquisa, and Argentina Agency CONICET.  CGGC also thanks the
Observatory of Paris in Meudon, that supported his stay to finish the
present work. The authors are in debt to A.  Magun and T. L\"uthi who
provided the calibrated data of the Bern patrol telescopes and of the
nulling interferometer at 89.4 GHz. They would also like to thank Dr. Lidia
van Driel-Gesztelyi for helpful discussions.
\end{acknowledgements}

\end{document}